\documentclass[apj]{emulateapj}

\usepackage{amsmath,amsthm,amssymb}

\usepackage{titlesec}

\usepackage{graphicx}
\usepackage{tabularx}

\usepackage{float}
\usepackage{color}
\usepackage{scalefnt}
\usepackage{hyperref}

\usepackage[numbib]{tocbibind}
\usepackage{enumitem}
\usepackage{multirow}

\usepackage[toc,page]{appendix}
\titleformat{\section}{\bfseries \large}{\thesection}{1em}{}
\titleformat{\subsection}{\bfseries \normalsize}{\thesubsection}{1em}{}
\titleformat{\subsubsection}{\itshape \normalsize}{\thesubsubsection}{1em}{}

\begin{document}

\title{The 3-Dimensional Architecture of the $\upsilon$ Andromedae Planetary System}
\shorttitle{The 3-D Architecture of $\upsilon$ Andromedae}
\author{Russell Deitrick}
\affil{Department of Astronomy, University of Washington, Seattle, WA 98195-1580, USA}
\email{deitrr@astro.washington.edu}
\author{Rory Barnes}
\affil{Department of Astronomy, University of Washington, Seattle, WA 98195-1580, USA}
\author{Barbara McArthur}
\affil{Department of Astronomy, University of Texas at Austin, TX 78712, USA}
\author{Thomas R. Quinn}
\affil{Department of Astronomy, University of Washington, Seattle, WA 98195-1580, USA}
\author{Rodrigo Luger}
\affil{Department of Astronomy, University of Washington, Seattle, WA 98195-1580, USA}
\author{Adrienne Antonsen}
\affil{Department of Astronomy, University of Washington, Seattle, WA 98195-1580, USA}
\and
\author{G. Fritz Benedict}
\affil{Department of Astronomy, University of Texas at Austin, TX 78712, USA}
\keywords{planetary systems, planets and satellites: dynamical evolution and stability, stars: individual (Upsilon Andromedae)} 
\date{}

\begin{abstract}

The Upsilon Andromedae system is the first exoplanetary system to have the relative inclination of two planets' orbital planes directly measured, and therefore offers our first window into the 3-dimensional configurations of planetary systems.  We present, for the first time, full 3-dimensional, dynamically stable configurations for the 3 planets of the system consistent with all observational constraints. While the outer 2 planets, c and d, are inclined by $\sim 30^{\circ}$, the inner planet's orbital plane has not been detected. We use N-body simulations to search for stable 3-planet configurations that are consistent with the combined radial velocity and astrometric solution. We find that only 10 trials out of 1000 are robustly stable on 100 Myr timescales, or $\sim 8$ billion orbits of planet b. Planet b's orbit must lie near the invariable plane of planets c and d, but can be either prograde or retrograde. These solutions predict b's mass is in the range $2$ - $9~M_{Jup}$ and has an inclination angle from the sky plane of less than $25^{\circ}$. Combined with brightness variations in the combined star/planet light curve (``phase curve''), our results imply that planet b's radius is $\sim 1.8~R_{Jup}$, relatively large for a planet of its age.  However, the eccentricity of b in several of our stable solutions reaches $> 0.1$, generating upwards of $10^{19}$ watts in the interior of the planet via tidal dissipation, possibly inflating the radius to an amount consistent with phase curve observations. 

\end{abstract}

\maketitle

\section{INTRODUCTION}

\subsection{Observations}
The $\upsilon$ Andromedae ($\upsilon$ And) planetary system was the first discovered multi-exoplanet system around a main sequence star and is possibly still the most studied multi-planet system other than our Solar System.  $\upsilon$ And b was discovered using the radial velocity (RV) technique by \cite{butler97} at Lick observatory.  Two years later, combined data from Lick and the Advanced Fiber-Optic Echelle spectrometer (AFOE) at Whipple Observatory revealed the presence of two additional planets, $\upsilon$ And c and d \citep{butler99}.  Follow-up by \cite{francois1999} confirmed that the existence of planets was the best explanation for the RV variations.  Even at the time of \cite{butler99}, the semi-major axes of the planets (0.059, 0.83, and 2.5 au), the minimum masses (0.71, 2.11, and 4.61 M$_{Jup}$), and eccentricities (0.034, 0.18, and 0.41) made it clear that this system was very unlike our Solar System, and it has presented a challenge to planet formation models that explain the Solar System. \cite{stepinski2000} confirmed the presence of these three RV signatures in the existing data using two different fitting algorithms, but stressed that the eccentricity of planet c was poorly constrained by the existing data.

Thus far, astrometry is one of the few techniques that can be used to break the $m \sin{i}$ degeneracy in the RV method for non-transiting planets \citep[the other is a relatively new technique that uses high-resolution spectra to directly observe the radial velocity of the planet; for example, see][]{brogi2012,rodler2012}. Astrometry is the process of measuring a star's movement on the plane of the sky, and hence provides 2-dimensional information which is orthogonal to RV. Because this measurement is made relative to other objects in the sky, it is extremely difficult to obtain the high precision necessary to detect planets. For small, close-in planets, the necessary precision is in the $\mu$as range \citep{quirrenbach2010}, since astrometry is more sensitive to planets with relatively large mass, low inclination and large semi-major axis. Nonetheless, \cite{mazeh99} reported a small, positive detection in the HIP data of an astrometric signal at the period of planet d, and derived an inclination of 156$^{\circ}$ and a mass of $10.1~M_{Jup}$. However, \cite{pourbaix2001} demonstrated that astrometric fits to the HIP data for 42 stars, including $\upsilon$ And, were not significantly improved by the inclusion of a planetary orbit, and that the inclinations for planets c and d could be statistically rejected. \cite{reffert2011} re-analyzed the HIP data and placed an upper mass limit on both planets c and d of $8.3 M_{Jup}$ and $14.2 M_{Jup}$, but did not claim true masses.

$\upsilon$ And became the first multi-planet system to have a positive astrometry detection above $3\sigma$ when \cite{mca2010} detected the orbits of planets c and d using \emph{Hubble Space Telescope} (\emph{HST}). Their orbital fit included all previously obtained RVs (including re-reduced Lick data), and added RVs from the Hobby-Eberly Telescope at McDonald Observatory. The combined astrometry+RV fit did not converge when planet b was included, indicating that planet b presents no astrometric signal. Indeed, using their Equation 8, which relates the RV and astrometry, planet b would be expected to have a signal of $\alpha \sim 40~\mu$as at an inclination of $\sim 3^{\circ}$, well below \emph{HST}'s detection limit of 0.25 mas. This non-detection puts a weak upper mass limit on planet b of $\sim~78 M_{Jup}$, as the planet would be astrometrically detectable by \emph{HST} at inclinations below $\sim 0.5^{\circ}$. 

\cite{mca2010} found that planets c and d have inclinations of $7.868^{\circ} \pm 1.003^{\circ}$ and $23.758^{\circ} \pm 1.316^{\circ}$, respectively, relative to the plane of the sky, with corresponding masses of $13.98^{+2.3}_{-5.3} M_{Jup}$ and $10.25^{+0.7}_{-3.3} M_{Jup}$. The mutual inclination between the two planets is $29.917^{\circ} \pm 1^{\circ}$. This value is quite unlike any mutual inclination found amongst the planets of our Solar System. Subsequent dynamical studies suggest that this may be the result of a three-body planet-planet scattering scenario in which one planet is ejected from the system \citep{barnes2011,libert2011}.

By looking at the infrared excess in the stellar spectrum attributed to planet b, \cite{harrington2006} attempted to chart the phase offset, i.e. the angle between the hottest point on the surface and the sub-stellar point. Assuming a radius of $< 1.4~R_{Jup}$, the observed amplitude of flux variation demanded that the planet's inclination must be $> 30^{\circ}$ from the plane of the sky.  Unfortunately, this study was based on only five epochs of data over a single orbit, and thus does not provide tight constraints. The infrared phase curve of $\upsilon$ And b was revisited with seven additional short epochs and one continuous $\sim 28$ hour observation by \cite{crossfield2010}.  The picture presented in \cite{crossfield2010} was consistent with \cite{harrington2006}, but they allowed larger radii in their model, finding that the inclination must be $> 28^{\circ}$ for a $1.3~R_{Jup}$ planet and $>14^{\circ}$ for a $1.8~R_{Jup}$.

There is marginal evidence for a fourth planet orbiting $\upsilon$ And. \cite{mca2010} found an improvement in their fit when a linear trend indicative of a longer period planet was included. Later, \cite{curiel2011} found a signal at 3848.9 days using the Lick \citep{fischer2003,wright2009} and ELODIE radial velocities \citep{naef2004}. These authors have taken this to be a fourth planet in the system, as suggested by \cite{mca2010}.  The \cite{mca2010} analysis used re-reduced Lick data (received by personal communication from Debra Fischer) for their combined RV and astrometry orbital fit. As explained in \cite{mcaHD128}, these data include updated $\gamma$ values (constant velocity offsets in the RVs) that removed this signal from the Lick data. Later, \cite{tuomi2011}
analyzed the older published RV data sets \citep{fischer2003,wright2009} and also found a period for this fourth planet (that was an artifact of the missing $\gamma$) of 2860 days, but noted that the data sets seemed inconsistent. \cite{tuomi2011} performed fits and calculated the Bayesian inadequacy criterion for the individual data sets. They found that the \cite{wright2009} Lick data produced a significantly different period for planet e (3860 days) and the Bayesian inadequacy criterion indicates that this data set has a $> 0.999$ probability of being inconsistent with the other data sets. While a longer period planet, indicated by a small slope in the
radial velocities, may exist in this system, the 4th planet signal reported by \cite{curiel2011} was a product of the earlier
reduction of the Lick data, which did not account for an instrument change that caused a shift in the $\gamma$. For this reason, we do not include this planet in our study.

The rotation and obliquity of $\upsilon$ And A are also of interest in this study (see section \ref{stability}). Measurements of $v \sin{i} = 9.6 \pm 0.5\text{ km s}^{-1}$ \citep{valenti2005} and stellar radius $R_{\star} = 1.64^{+0.04}_{-0.05}~R_{\odot}$ \citep{takeda2007} limit the rotation period to be $\lesssim 8$ days for physical values of $i$, however the only measured period consistent with these data is $7.3$ days \citep{simpson2010}. This period suggests an obliquity $i \sim 60^{\circ}$ (measured from the sky plane), but the signal at this period is very weak and it is impossible to distinguish this obliquity from $i \sim 120^{\circ}$ (spinning in the opposite sense). 

\subsection{Theory}
\label{introtheory}
Numerous dynamical studies of $\upsilon$ Andromedae have been performed, both numerical and analytical.  Early studies focused on the stability of the system using N-body models and analytic theory, prior to the astrometric detection of planets c and d \citep{mca2010}.  These studies showed that the stability of the system is highly sensitive to the eccentricities of planets (d in particular) \citep{laughlin1999,bq2001,bq2004}, the relative inclinations of the planets \citep{rivera2000,stepinski2000,chiang2001,lissauer2001,ford2005,michtchenko2006}, their true masses \citep[since only minimum masses were known prior to][]{mca2010} \citep{rivera2000,stepinski2000,ito2001}, the effect of general relativity on planet b's eccentricity \citep{nagasawa2005,adams2006,migaszewski2009}, and the accuracy of the RV data \citep{lissauer1999,stepinski2000,gozdziewski2001}, and also found that there were stable regions only between planets b and c and exterior to planet d \citep{rivera2000,lissauer2001,barnes2004,rivera2007}. The dynamical study of \cite{mca2010} found stable configurations for all three planets on a timescale of $10^5$ years, and constrained the inclination of planet b to $i < 60^{\circ}$ or $i > 135^{\circ}$.  It has also been demonstrated that analytical or semi-analytical theory does not adequately describe the dynamics of the system \citep{veras2007}, unless taken to very high order in the eccentricities \citep{libert2007}, though these studies assumed coplanarity since the inclinations of the planets were not known at the time.

Other studies dealt with the evolution and formation of certain features of the system, in particular the apparent alignment of the pericenters (or ``apsidal alignment'', $\Delta \varpi$, noted by \cite{rivera2000,chiang2001}) and large eccentricities of planets c and d. Some have investigated whether the present day system could have been produced by interactions with a dissipating disk \citep{chiangmurray2002,nagasawa2003}, by planet-planet scattering \citep{nagasawa2003,ford2005,barnes2007,ford2008,barnes2011}, by interactions with the stellar companion, $\upsilon$ And B \citep{barnes2011}, by secular or resonant orbital evolution \citep{jiang2001,malhotra2002,ford2008,libert2011}, or by accelerations acting on the host star \citep{namouni2005}. \cite{michtchenko2004} found that $\Delta \varpi$ can be in a state of circulation, libration or a ``non-linear secular resonance'', all within the observational uncertainty. In short, the dynamical evolution of the system appears to be highly sensitive to the initial conditions, and planet-planet scattering appears to be the most promising explanation for its current state.

In particular, \cite{mca2010} noted that the true masses of planets c and d naturally resolve a difficulty in explaining the system's formation. When only the minimum masses of the two planets were known, it was generally assumed in dynamical analysis that planet d was the larger ($m \sin{i_c} = 1.8898~M_{Jup}$ and $m \sin{i_d} = 4.1754~M_{Jup}$), however, mechanisms that can excite the eccentricities of the planets, such as resonance crossing \citep{chiang2002} or close encounters \citep{ford2001}, tend to result in the smaller mass planet having the larger eccentricity. Some authors noted that because planet d is observed to have the larger eccentricity ($e_c =  0.245\pm 0.006, e_d = 0.316\pm0.006$ \citep{mca2010}), the formation of the system may have required the presence of a gas disk or the ejection of an additional low-mass planet \citep{chiang2002, rivera2000, ford2005, bg2007}. 

Finally, \cite{burrows2008} produced theoretical pressure-temperature profiles and spectra for several close-in giant planets, including $\upsilon$ And b, and compared these results with the phase curve in \cite{harrington2006}.  Unfortunately, the phase curve data were too sparse to make any conclusions about the size or structure of the atmosphere or the planet's inclination, only that a range of radii and inclinations are consistent with the data, and that a temperature inversion in the atmosphere is also consistent.  The authors suggested that more frequent observations and multi-wavelength data may break the degeneracies in their model.

\subsection{This Work}
Observations account for the inclinations and true masses of planets c and d, but the inclination and true mass of planet b remain undetermined.  Early dynamical studies found stable regions of parameter space for the coplanar, three planet system, however, stable configurations have not previously been identified for three planets over long timescales since the large mutual inclination between planets c and d was discovered by astrometry.  Additionally, it is unclear whether the phase curve observations \citep{harrington2006,crossfield2010} are consistent with the RV+astrometry observations \citep{mca2010}.  Here we explore all the above issues using dynamical models.

This work is a sweep through parameter space for stable configurations for all three planets, following up on the dynamical analysis in \cite{mca2010}. An acceptable configuration in this study is one that 1) satisfies the RV+astrometry fit \citep{mca2010}, 2) is dynamically stable, and 3) is consistent with the IR phase curve measurements \citep{crossfield2010}. Our results satisfy all three criteria, however, reconciliation with the phase curve measurements requires planet b to have an inflated radius and motivates us to include tidal heating in this study.

Section 2 describes the methods that we use to explore parameter space, model the dynamics, and estimate tidal heating in planet b. Section 3 describes the results we obtain for stability and system evolution. Section 4 focuses on tidal heating and reconciliation with \cite{crossfield2010}. In Section 5, we discuss our results in the context if previous studies, and summarize our conclusions in Section 6.

\section{METHODS}
\subsection{Updated Orbits and Parameter Space}

As in \cite{mca2010}, all RV data sets are re-examined using the published errors (which are large for the AFOE data, in particular). The Lick data set has been re-reduced since the original publications, resulting in new $\gamma$ offsets. This updated set, published recently in \cite{fischer2014}, was used in \cite{mca2010} (received by personal correspondance with D. Fischer), and similarly we use it here. The RV fit is performed on each data set individually and compared with the other data sets for consistency, and large outliers are examined and removed, if necessary.  

Trials are generated by drawing randomly from within the uncertainties of the $\chi^2$ best fit to the RV and astrometric data. Table \ref{tab:params_tab} shows the parameter space explored. The astrometric constraints on the semi-major axes of planets c and d are ignored as the RV derived periods provide much stricter constraints on these quantities. Trials are not generated taking into account interdependencies of the various parameters, however, because of the small size of the uncertainties all trials should be faithful to the data. Nevertheless, we include $\chi^2$ values for our stable cases to confirm consistency. This work is not meant to represent a complete analysis of all possible configurations. We are merely establishing the existence of stable cases within the observational constraints

The nominal eccentricity of planet d is $0.316$, however, we find that very few trials are stable above $e_d \sim 0.3$ (see Figure \ref{fig:hist1}), so we apply a much looser constraint to the lower bound of planet d's eccentricity, drawing from a uniform distribution across the domain $0.246 < e_d < 0.322$, rather than a Gaussian distribution.

\begin{table*}
\centering
\caption{\textbf{Parameter space:} parameters are drawn from Gaussian distribution with center (standard deviation) except where noted} 
\newcolumntype{R}{>{\raggedleft\arraybackslash}X}

\begin{tabularx}{\textwidth}{lRRR}
\hline\hline \\ [-1.5ex]
 & $\upsilon$ And b & $\upsilon$ And c & $\upsilon$ And d \\ [0.5ex]
\hline \\ [-1.5ex]

$e$ & 0.01186 (0.006) & 0.2445 (0.1) & 0.316 (0.07/0.006)* \\
$i$ ($^{\circ}$) & 90 (90)$^{\dagger}$ & 11.347 (3.0) & 25.609 (3.0) \\
$\omega$ ($^{\circ}$) & 44.519 (24.0) & 247.629 (2.2) & 252.991 (1.32) \\
$\Omega$ ($^{\circ}$) & 180 (180)$^{\dagger}$ & 248.181 (8.5) & 11.425 (3.31) \\
$P$ (days) & 4.61711 (0.00018) & 240.937 (0.06) & 1281.439 (2.0) \\
$T$ (days) & 2450034.058 (0.3) & 2449922.548 (1.5) & 2450059.072 (4.32) \\
$K$ (m/s) & 70.519 (0.368) & 53.4980 (0.55) & 67.70 (0.55) \\

\hline \\ [-1.5ex]

\end{tabularx}

\small
*drawn from a uniform distribution with lower/upper bounds shown \\
$^{\dagger}$uniform distribution \\
\label{tab:params_tab}
\end{table*}

\subsection{N-Body}
For the stability analysis, we use \emph{HNBody} \citep{rauch2002}, which contains a symplectic integrator for central-body-like systems, i.e., systems in which the total mass is dominated by a single object. This symplectic scheme alternates between Keplerian motion and Newtonian perturbations at each timestep \citep[see][]{wisdom1991}. During one half-step (the ``kick'' step), all gravitational interactions are calculated  and the momenta are updated accordingly. During the other half-step (the ``drift'' step), the system is advanced along Keplerian (2-body) motion, using Gauss's $f$ and $g$ equations \citep[see][]{danby}. The entire integration is done in Cartesian coordinates. Unlike \emph{Mercury} \citep{chambers1999}, post-Newtonian (general relativisitic) corrections are included as an optional parameter in \emph{HNBody}, and we utilize them here.

While \emph{HNBody} is fast and its results compare well with results from \emph{Mercury}, its definitions of the osculating elements used at input and output differ slightly from \emph{Mercury}'s. The mass factor used in the definitions of semi-major axis and eccentricity (for astrocentric or barycentric elements) does not include the planet's mass; in other words, the planet is treated as a zero mass particle during input and output conversions between Cartesian coordinates and osculating elements. Because of this, the use of osculating elements during input can result in incorrect periods and Cartesian velocities. For most planetary systems, which have poorer constraints on the periods and semi-major axes of the planets, this aspect makes little difference, but for $\upsilon$ Andromedae the periods are known with the high precision that comes with $> 15$ years of RVs, and so all of our input and output from \emph{HNBody} is done in Cartesian positions and velocities, to ensure proper orbital frequencies.

A second important conversion must be done to account for a difference in units. \emph{HNBody} and \emph{Mercury} enforce the relationship $GM_{\odot}D^2/au^3 = k^2$, where $k$ is the Gaussian gravitational constant (defined to be  $0.01720209895$ au$^{3/2}~M_{\odot}^{-1/2}~D^{-1}$), $D$ is the length of the day, and au is the astronomical unit, based on the IAU defitions prior to 2012. The current accepted IAU units fix the au to be exactly 149,597,870,700 m and the constant $k$ is no longer taken to be a constant value. This redefinition was done for ease of use and to allow for reconciliation with the length-contraction and time-dilation of Einstein's relativity. Because these N-body models were developed prior to this redefinition, $k$ was used as a fundamental constant, as that allowed for better accuracy in Solar System integrations when the au was uncertain. We believe these integrators still accurately represent the dynamics of planetary systems, since these models do not attempt to account for length-contraction and time-dilation and in any case these effects will be very small. Note that HNBody does account for relativistic precession, which is important for the motion of planet b, as mentioned in Section \ref{introtheory}.

The units utilized by the integrator are then au, days, and solar masses. For observations of exoplanets, SI units are more sensible, but the values of $G$, $M_{\odot}$, and au need not obey the above constraint. Hence the simulated orbital periods will not be equal to the measured orbital periods unless we first convert from SI units, which do not obey this constraint, to IAU units, which do. We accomplish this by choosing a value for the au which satisfies $GM_{\odot}D^2/au^3 = k^2$, given the $G$, $M_{\odot}$, and $D$ used in the model of the observations. We then check that this definition of the au correctly reproduces the orbital periods of the planets when calculated using $k$ and Kepler's third law,
\begin{equation}
T^2 = \frac{4 \pi^2}{k^2(M_{\star}+m_p)} a^3,
\end{equation}
where $T$ is the planet's period in days, $M_{\star}$ and $m_p$ are the masses of the star and planet in solar masses, and $a$ is the planet's semi-major axis in au. Finally, to verify that \emph{HNBody} ``sees'' the correct periods, we run 2-body integrations for each planet at high time resolution and performed FFTs on the planets' velocities, confirming that this approach is the most accurate.

For the reasons described above, we take the orbital elements from the RV+astrometry, convert to Cartesian coordinates for dynamical modelling, then convert back to orbital elements for stability analysis. For analysis of the orbital evolution, we convert from Cartesian ``line-of-sight'' coordinates to Cartesian invariable plane (the plane perpendicular to the total angular momentum of the system) coordinates prior to the conversion back to orbital elements. The invariable plane of the system is inclined by $\sim 15^{\circ}$ from the sky, so inclinations measured from this plane are similar to those measured from the sky-plane. See the appendix for a potential pitfall in modelling astrometrically measured orbits.

The initial parameter space includes inclinations of planet b from $0^{\circ}$ to $180^{\circ}$, which is to say that any inclination is consistent with the observations. Extremely low inclination, $i_b < 1^{\circ}$ or $i_b > 179^{\circ}$, places the mass of planet b in the brown dwarf range. Inclinations $\gtrsim 90^{\circ}$ are retrograde orbits with respect to the orbits of the outer two planets (note that the outer system's invariable plane is inclined only $\sim 10^{\circ}$ from the sky plane). The observations allow for such configurations, and we cannot rule them out based on dynamical stability.

\subsection{Tidal Theory}

We use a constant phase-lag (CPL) model \citep{darwin1880} to estimate the amount of tidal energy that could be dissipated in the interior of $\upsilon$ And b. This model is described in detail in appendix E of \cite{barnestidalvenus} \citep[see also][]{ferrazmello2008}. In short, the model treats the tidal distortion of the planet as a superposition of spherical harmonics with different frequencies, which sum to create a tidal bulge that lags the rotation by a constant phase. The strength of tidal effects is contained in the parameter $Q$, called the tidal quality factor, estimated to be $\gtrsim 10^4$ for the planet Jupiter \citep{goldreichsoter66,yoderpeale81,aksnes2001}. For close in Jovians like $\upsilon$ And b, \cite{ogilvie2004} suggest $Q$ could be as high as $5 \times 10^7$, hence we model a range of $Q$ values from $10^4 - 10^8$. 

The CPL model, commonly used in the planetary science community, is only an approximate representation of tidal evolution \citep[for an in-depth discussion of the limitations of the model, see][]{efroimsky2013}. However, at the eccentricities we explore here ($e \lesssim 0.15$), results from this model are qualitively similar to other tidal models. The CPL model has the advantage of fast computation and is accurate enough for our purpose here, which is merely to establish the possibility of tidal heating in the interior of planet b. Given the uncertainty in the model and in the properties of the planet, our results should not be taken to be a precise calculation of the planet's tidal conditions.

\section{ORBITAL DYNAMICS}

In this section, we present our results regarding the dynamics of the system. In Section \ref{stability}, we examine the stability of the system. In Section \ref{systemevol}, we compare the orbital evolution in our favored cases.

\subsection{Stability}
\label{stability}
\begin{figure*}[t]

\includegraphics[width=6.5in]{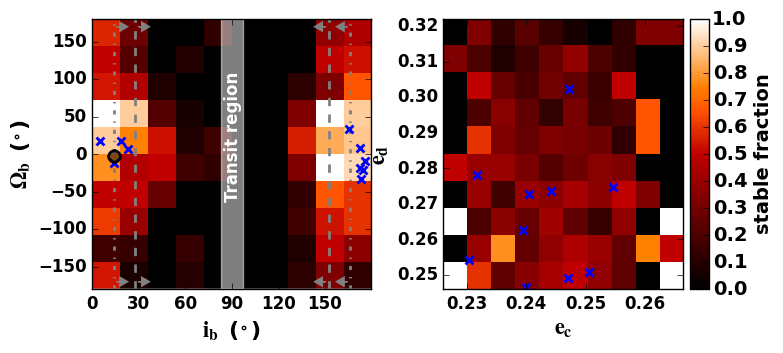}
\caption{Stable fraction ($f_s = N_s/N$, where $N_s$ is the number of trials that survived 1 Myr in each bin and $N$ is the total number of trials in each bin), for different parameters. Left: Longitude of ascending node ($\Omega_b$) vs. inclination ($i_b$) of planet b. The \emph{x}'s represent our robustly stable cases that survived for 100 Myr with no signs of chaos. The black circle represents the average fundamental plane of a system with planets c and d only. Shown also are the inclinations predicted by \cite{crossfield2010} (dashed line: $R_p > 1.3 R_{Jup}$; dotted line: $R_p > 1.8 R_{Jup}$) and the region for which the planet would transit the host star. Higher stability occurs at $i_b \lesssim 40^{\circ}$ and $i_b \gtrsim 140^{\circ}$. Right: Eccentricity of planet d vs. eccentricity of planet c. Stability is most dependent on $e_d$, which must remain $\lesssim 0.3$ for the system to remain stable. Stability seems uncorrelated with $e_c$ (the bright colored bins on the far left and far right contain only 1-2 trials each, and are therefore not necessarily regions of high stability).}
\label{fig:hist1}
\end{figure*}

We run our initial set of 1,000 trials for 1 Myr and flag as unstable those trials in which one or more planets were lost. The resulting stability maps are shown in Figure \ref{fig:hist1}. Here we show the most relevant parameters, i.e., those with the largest uncertainty ($\Omega_b$ and $i_b$) and those which have a large effect on stability ($e_c$ and $e_d$). The coordinate system used here is the RV+astrometry (``line-of-sight'') coordinate system, in which $i$ is measured from the plane of the sky and $\Omega$ is measured counterclockwise from North.

We note regions of greater stability concentrate around inclinations for planet b of $\lesssim 40^{\circ}$ and $\gtrsim 140^{\circ}$ and ascending nodes of $\sim 0^{\circ}$. A chasm of instability lies across inclinations of $\sim 60^{\circ}$ to $\sim 140^{\circ}$.

Next we examine the eccentricity and inclination evolutions for all trials in which three planets survived 1 Myr. Many of these ``stable'' cases exhibit chaotic evolution, with planet b even reaching eccentricities of $\sim 0.9$. We assume trials with chaotic evolution are in the process of destabilizing and should be discarded. This leaves us with $\sim 30$ trials which we then ran for 100 Myrs. Trials in which all planets survive 100 Myr integrations with no chaotic evolution or systematic regime changes are considered robustly stable. These 10 cases are plotted as x's in Figure \ref{fig:hist1}.

$\upsilon$ Andromedae is estimated to be $\sim 3$ Gyrs old \citep{takeda2007}; our ideal goal would thus be to demonstrate stability over this timespan; however, because a very small timestep is necessary to resolve the 4.6 day orbit of planet b, simulations of the system on spans of Gyrs are computationally prohibitive. Hence, we limit ourselves to a domain of 100 Myrs ($1/30$th of the system's lifetime), and note that this length of time corresponds to $\sim8$ billion orbits of planet b and that no previous study has been able to show stability for all three planets on this timescale.  

The $\chi^2$ results for our stable cases are shown in Table \ref{chisq_tab}. Here, $\chi^2$ represents the goodness of fit of each configuration to the data, and would ideally be equal to the number of degrees of freedom (DOF) in the model of the data. Configuration {\bf{PRO1}} is chosen as our nominal case because it has the lowest $\chi^2$ value of the prograde (orbit of planet b) cases. 

\begin{figure*}

\includegraphics[width=3.25in]{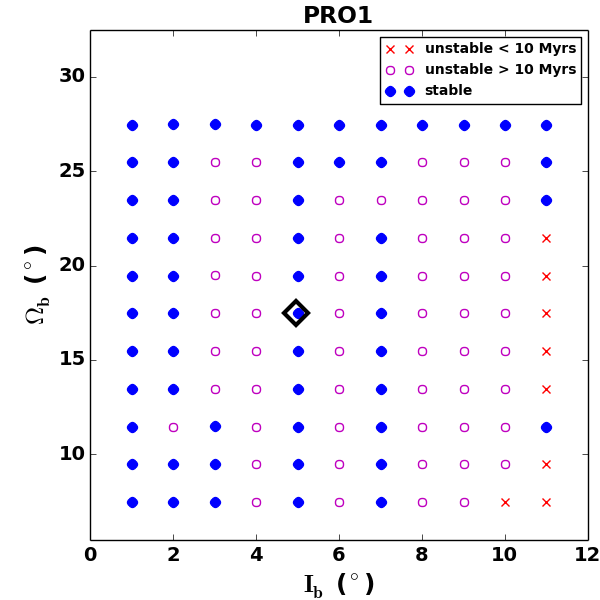}
\includegraphics[width=3.25in]{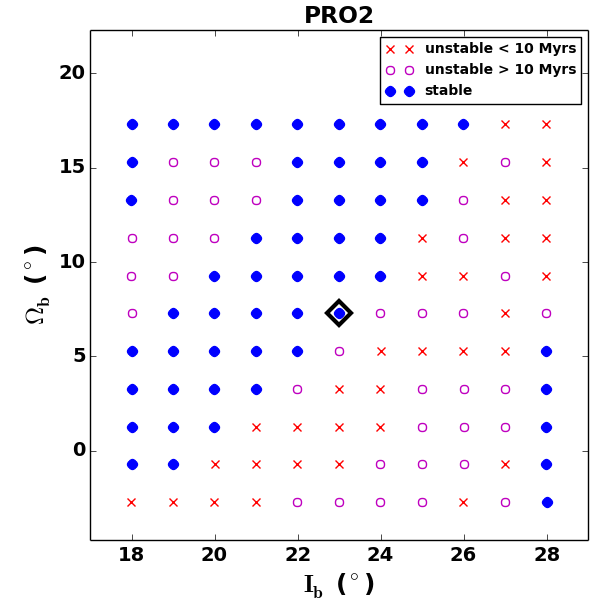} \\
\includegraphics[width=3.25in]{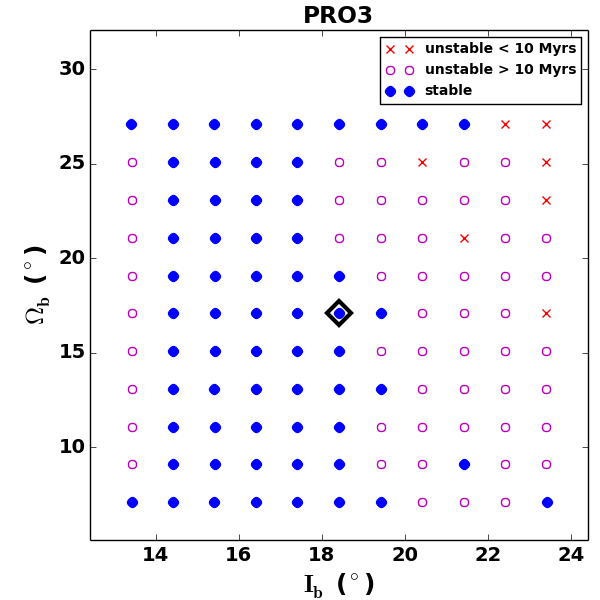}
\includegraphics[width=3.25in]{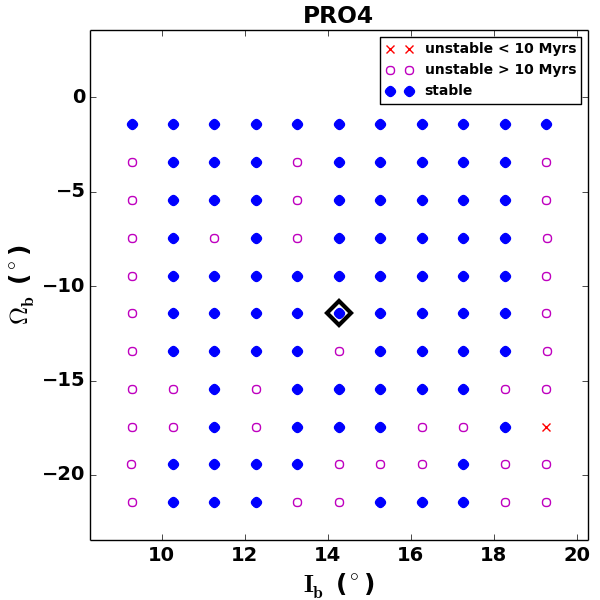}
\caption{Stable regions surrounding our prograde trials, varying the orbital plane of planet b. Red \emph{x}'s are trials that had a planet ejected in less than 10 Myrs, pink circles displayed chaotic evolution but no ejections over 10 Myrs, and blue solid circles are those which are truly stable over 10 Myrs. The original trials are surrounded by the black diamonds.}
\label{fig:pro1}
\end{figure*}

For our 4 prograde, stable trials, we generate the stability maps shown in Figure \ref{fig:pro1}.  Keeping all other parameters constant, we vary the inclination and ascending node of planet b to further explore these ``islands'' of stability. In order to keep all our cases consistent with the observations of planet b, we adjust its mass with changes in inclination and subsequently must adjust its semi-major axis to maintain the observed period. Thus changes in inclination imply not only a different mass via the $m\sin{i}$ degeneracy, but also a change in semi-major axis.

In all cases we see that our solutions occupy stable regions of phase space. {\bf PRO2} and {\bf PRO3} appear to be perched on the edges of two large stable regions, while {\bf PRO1} occupies a very narrow stable ``inclination stripe'' at $\sim 5^{\circ}$ and $\sim 2-3^{\circ}$, respectively. As in Figure \ref{fig:hist1}, inclination in Figure \ref{fig:pro1} is measured from the sky-plane, not the invariable plane of the system. 

\begin{table}[h]
\centering
\caption{\textbf{Stable configurations}}

\begin{tabular}{lr}
\hline\hline \\ [-1.5ex]
Trial & $\chi^2$ (DOF = 811) \\
\hline \\ [-1.5ex]

\bf{PRO1} & 779 \\ %83
\bf{PRO2} & 2218  \\ %973
\bf{PRO3} & 2353 \\ %1749
\bf{PRO4} & 3378  \\ %2629
\bf{RETRO1} & 672 \\ %1581
\bf{RETRO2} & 725  \\ %10
\bf{RETRO3} & 1292  \\ %1005
\bf{RETRO4} & 1524  \\ %1533
\bf{RETRO5} & 1917  \\ %1128
\bf{RETRO6} & 3062  \\ %700

\hline 
\end{tabular}
\label{chisq_tab}
\end{table}

\begin{table*}[h]
\centering
\caption{\textbf{Orbital parameters for stable, prograde trials}}

\begin{tabular}{lrrrrcrrrr}
\hline\hline \\ [-1.5ex]
%case 83
ID & planet & $m$ (M$_{Jup}$) & $P$ (days) & $a$ (au) & $e$~ & $i$ ($^{\circ}$) & $\omega$ ($^{\circ}$) & $\Omega$ ($^{\circ}$) & $MA$ ($^{\circ}$)\\ [0.5ex]
\hline \\ [-1.5ex]

{\bf PRO1} & b &  8.02 & 4.61694 & 0.059496 & 0.003547 & 4.97 & 48.39 & 17.47 & 129.43 \\
{\bf PRO1} & c &  8.69 & 240.92 & 0.830939 & 0.254632 & 12.62 & 245.89 & 259.40 & 153.03 \\
{\bf PRO1} & d & 10.05 & 1281.08 & 2.532293 & 0.274677 & 24.55 & 253.71 & 10.22 & 83.16 \\

\hline \\ [-1.5ex]
%case 973

{\bf PRO2} & b & 1.78 & 4.61716 & 0.059408 & 0.011769 & 22.99 & 51.14  & 7.28 & 103.53\\
{\bf PRO2} & c & 10.78 & 241.02 & 0.831580 & 0.247042 & 10.09 & 248.74 & 256.34 & 154.47\\
{\bf PRO2} & d & 8.86 & 1282.57 & 2.533539 & 0.249090 & 28.30 & 253.27 & 9.97 & 82.57\\

\hline \\ [-1.5ex]
%case 1749

{\bf PRO3} & b & 2.20 & 4.61726 & 0.059415 & 0.003972 & 18.41 & 44.98 & 17.09 & 148.28\\
{\bf PRO3} & c & 8.92 & 240.91 & 0.830954 & 0.247205 & 12.36 & 247.69 & 243.16 & 155.52\\
{\bf PRO3} & d & 9.92 & 1281.41 & 2.532645 & 0.302355 & 24.78 & 252.60 & 9.59 & 83.20 \\

\hline \\ [-1.5ex]
%case 2629

{\bf PRO4} & b & 2.81 & 4.61693 & 0.059421 & 0.021686 & 14.27 & 49.50 & 348.56 & 150.17\\
{\bf PRO4} & c & 9.01 & 240.93 & 0.831030 & 0.231669 & 12.31 & 244.39 & 243.47 & 154.06\\
{\bf PRO4} & d & 9.98 & 1280.56 & 2.531579 & 0.278130 & 24.74 & 252.37 & 9.90 & 83.77 \\

\hline 
\end{tabular}
\label{orbits_tab}
\end{table*}

An additional complication to the dynamics of the system is the oblateness of the host star.  \cite{migaszewski2009} showed the importance of $J_2$ (the leading term in the gravitational quadrupole moment) of the star in their secular analysis, though they used a stellar radius of $R = 1.26~R_{\odot}$, significantly smaller than the current best measurement of $R = 1.631~\pm~0.014~R_{\odot}$ \citep{baines2008}. To verify the importance of oblateness, we simulated our best prograde $\chi^2$ case ({\bf PRO1}), varying $J_2$ from $10^{-5}$ to $10^{-2}$ and $R_{\star}$ from $1.26~R_{\odot}$ to $1.63~R_{\odot}$.  We find that values of $J_2\gtrsim10^{-3}$ cause the system to become unstable, and lower values significantly change the eccentricity evolution of planet b. Unfortunately, the $J_2$ value of the star is not known and there exists some disagreement regarding its radius, and thus a detailed exploration of parameter space including these two additional parameters is beyond the scope of this work. Therefore, in our primary analysis, the quadrupole moment is ignored ($J_2 = 0$). 

\subsection{System Evolution}
\label{systemevol}

\begin{figure*}
\includegraphics[width=3.25in]{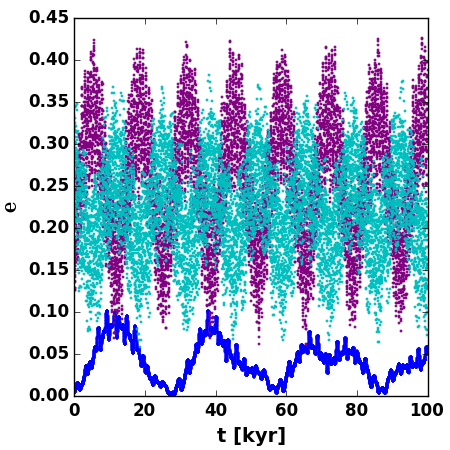}
\includegraphics[width=3.25in]{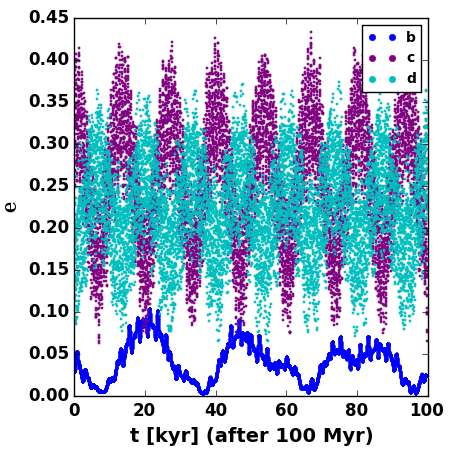}\\
\includegraphics[width=3.25in]{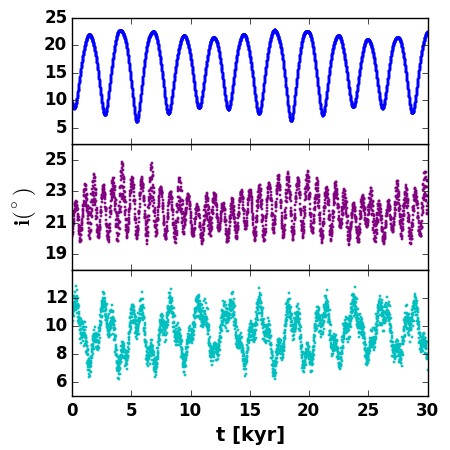}
\includegraphics[width=3.25in]{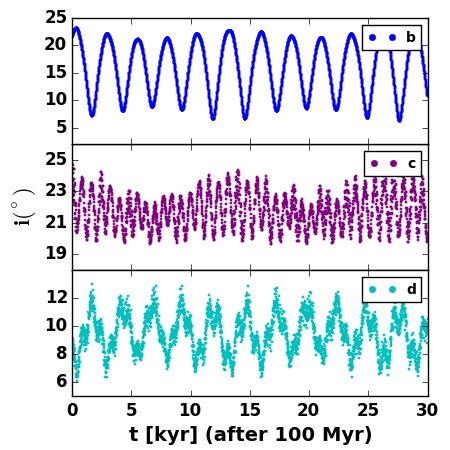}
\caption{Eccentricity evolution (top panels) and inclination evolution (bottom panels) for planet b (blue), planet c (purple), and planet d (cyan) over 100,000 years, from the current epoch (left) and after a 100 Myr integration (right), in the {\bf PRO1} system. The eccentricity evolution of planet b may appear unstable, but as seen in the right panel, the pattern is periodic over at least 100 Myr timescales. Inclinations here are measured from the invariable plane of the system, rather than the sky-plane.}
\label{fig:evol1}
\end{figure*}

\begin{figure*}
\includegraphics[width=3.25in]{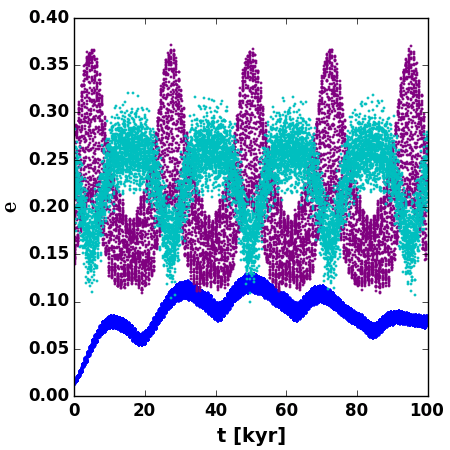}
\includegraphics[width=3.25in]{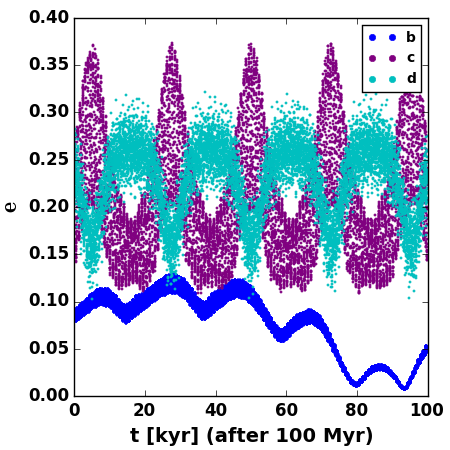}\\
\includegraphics[width=3.25in]{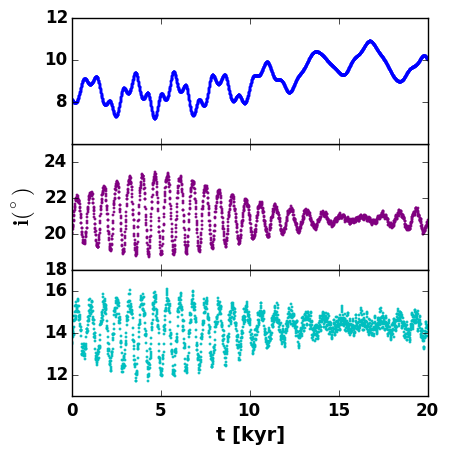}
\includegraphics[width=3.25in]{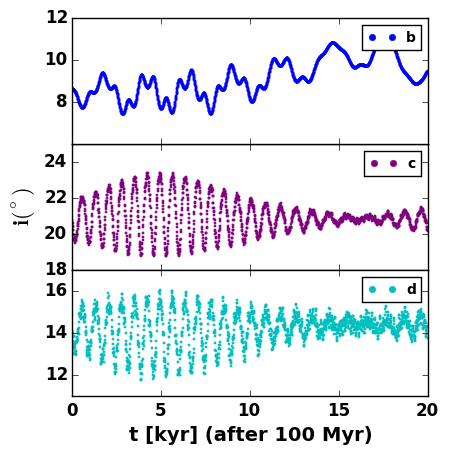}
\caption{As in Figure \ref{fig:evol1} but for case {\bf PRO2}}
\label{fig:evol2}
\end{figure*}

\begin{figure*}
\includegraphics[width=3.25in]{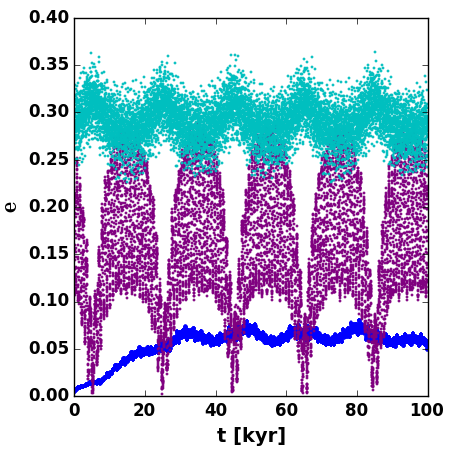}
\includegraphics[width=3.25in]{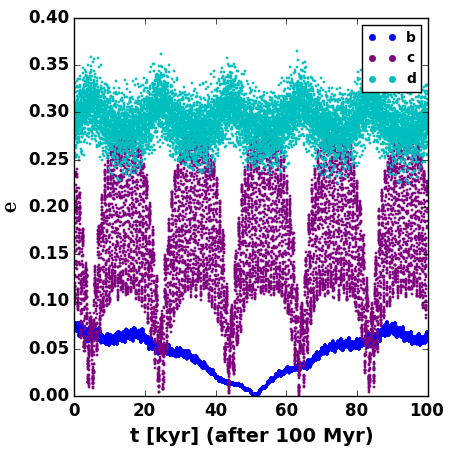}\\
\includegraphics[width=3.25in]{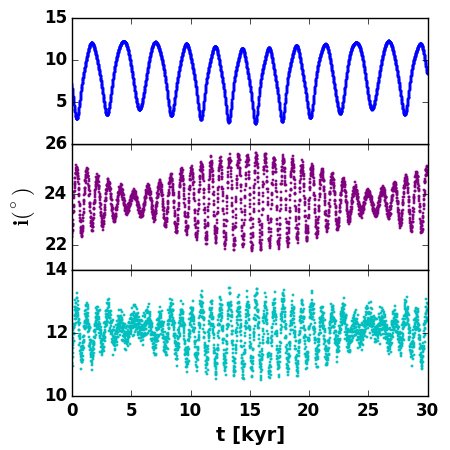}
\includegraphics[width=3.25in]{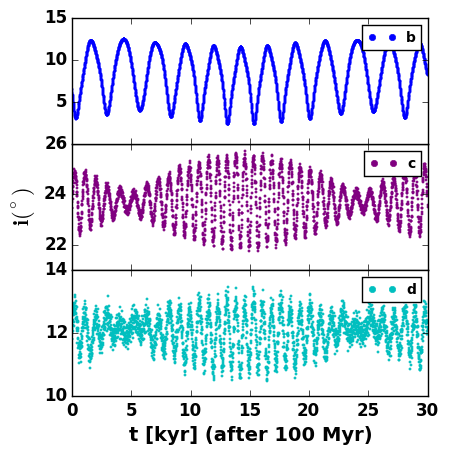}
\caption{As in Figure \ref{fig:evol1} but for case {\bf PRO3}}
\label{fig:evol3}
\end{figure*}

\begin{figure*}
\includegraphics[width=3.25in]{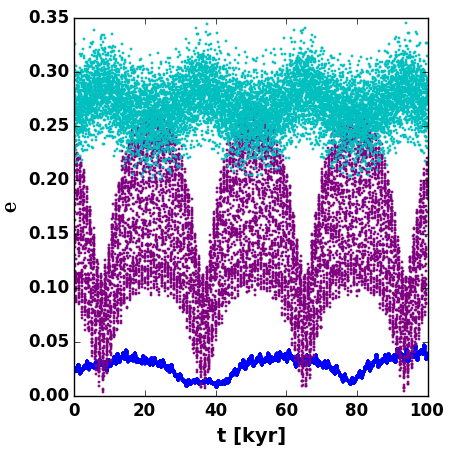}
\includegraphics[width=3.25in]{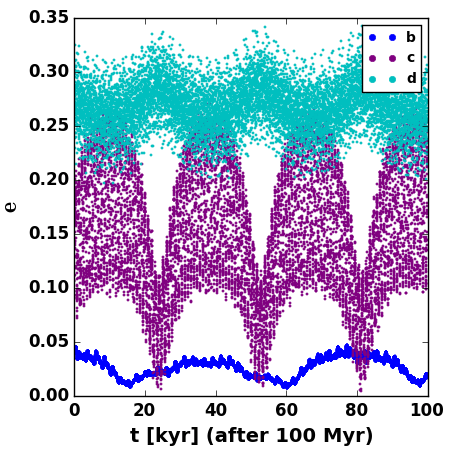}\\
\includegraphics[width=3.25in]{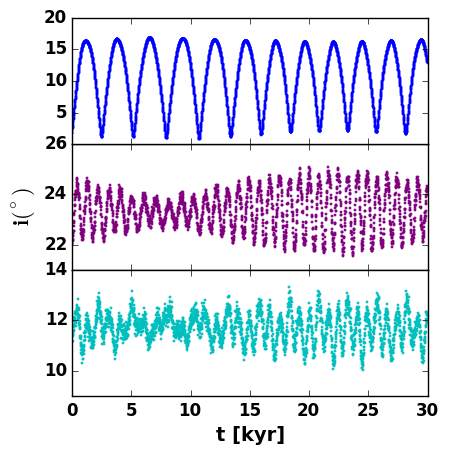}
\includegraphics[width=3.25in]{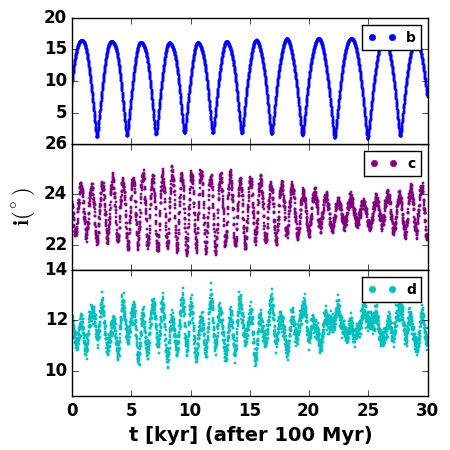}
\caption{As in Figure \ref{fig:evol1} but for case {\bf PRO4}}
\label{fig:evol4}
\end{figure*}

The eccentricity and inclination evolutions for our prograde trials are shown in Figures \ref{fig:evol1} - \ref{fig:evol4}. In these figures, inclination is measured from the invariable plane, that is, a plane perpendicular to the total angular momentum vector of the system. We see that for all cases, the evolution of the eccentricities and inclinations is periodic for at least 100 Myrs ($\sim 8$ billion orbits of planet b). At a glance, one might expect for {\bf PRO1} to lose planet b, however, as shown in Figure \ref{fig:evol1}, the pattern seen in its eccentricity evolution is repeated reliably over many orbits, and therefore the configuration is robustly stable.

\begin{figure*}
\includegraphics[width=3.25in]{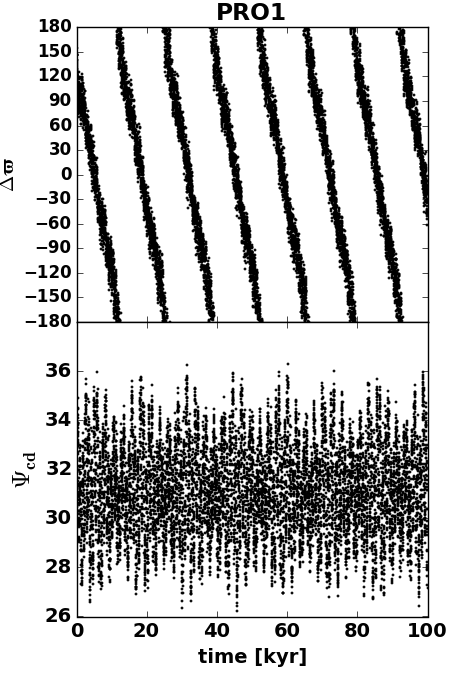}
\includegraphics[width=3.25in]{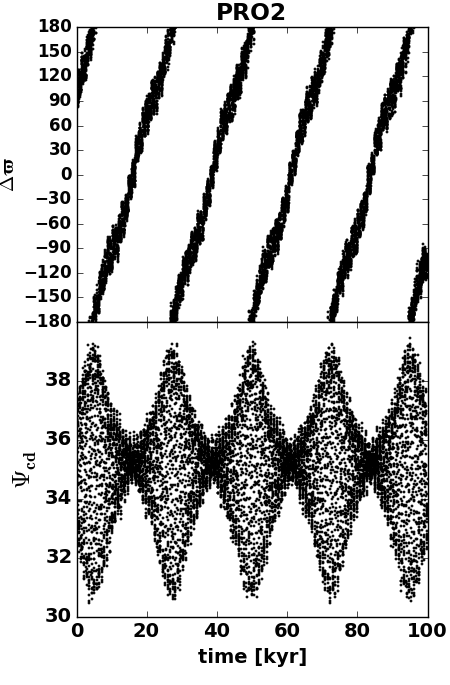}
\caption{Secular behavior of planets c and d. $\Delta \varpi = \varpi_d - \varpi_c$ circulates in both cases. The mutual inclination, $\Psi_{cd}$, oscillates about $\sim 31^{\circ}$ in {\bf PRO1} and $\sim 35^{\circ}$ in {\bf PRO2} with a $\sim 10^{\circ}$ amplitude in both.}
\label{fig:secular1}
\end{figure*}

\begin{figure*}
\includegraphics[width=3.25in]{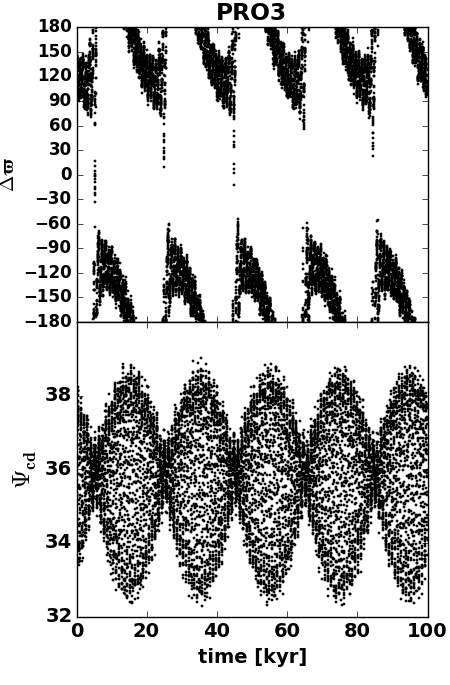}
\includegraphics[width=3.25in]{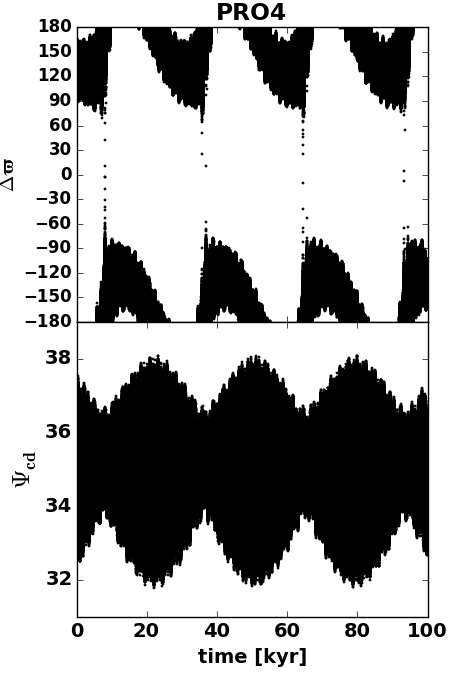}
\caption{Secular behavior of planets c and d. $\Delta \varpi = \varpi_d - \varpi_c$ librates with intermittent circulation in both cases. {\bf PRO4} is plotted with finer time resolution to show the circulation, which occurs very quickly. The mutual inclination, $\Psi_{cd}$, oscillates about $\sim 35^{\circ}$ in both cases with a $\sim 6^{\circ}$ amplitude.}
\label{fig:secular2}
\end{figure*}

For planets c and d, Figures \ref{fig:secular1} and \ref{fig:secular2} show $\Delta \varpi = \varpi_d - \varpi_c$ (the difference of the longitudes of pericenter) and their mutual inclination $\Psi_{cd}$ for our cases in which planet b is in prograde motion. We see that $\Delta \varpi$ undergoes circulation in cases {\bf{PRO1}} and {\bf PRO2}, and librates around anti-alignment in cases {\bf PRO3} and {\bf PRO4}, although it is very close to the separatrix in both. The amplitudes of libration are $\sim 240^{\circ}$ and $\sim 210^{\circ}$, for {\bf PRO3} and {\bf PRO4}, respectively, and RMS values about the libration center ($180^{\circ}$) are $55^{\circ}$ and $47^{\circ}$, respectively. The mutual inclination between planets c and d oscillates between $\sim 30^{\circ}$ and $\sim 40^{\circ}$ in cases {\bf PRO2}, {\bf PRO3}, and {\bf PRO4}. The angle explores a slightly wider range in case {\bf PRO1}.

\section{TIDAL HEATING}

The phase curve measurements \citep{crossfield2010} require planet b to have a radius of $1.3~R_{Jup}$ at an inclination of $28^{\circ}$ ($1.8~R_{Jup}$ at $i = 14^{\circ}$). This large radius could be explained by a combination of intense stellar irradiation and tidal heating in the planet's interior. To that end, we present predictions of tidal energy dissipation in several of our cases.

In reference to the planet HD 209458 b, \cite{ibgui2009} found that early episodes of high eccentricity can cause tidal dissipation of $\sim 10^{19}$ watts of power, which helps to explain the planet's radius of $1.3~R_{Jup}$, though it may not be necessary, considering the stellar flux received. We find that planet b in our prograde cases experiences significant eccentricity evolution (Figures \ref{fig:evol1} - \ref{fig:evol4}), which should trigger similar episodes of tidal heating. As discussed in section \ref{discuss}, this may be necessary to reconcile our results with that of \cite{crossfield2010}.  

\begin{figure*}
\includegraphics[width=6.5in]{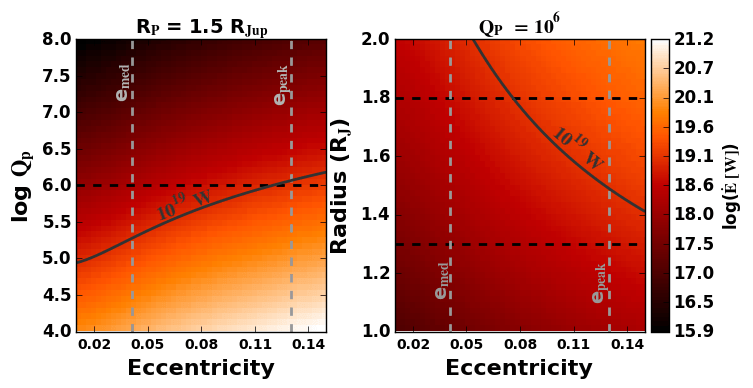}

\caption{Tidal heating for $\upsilon$ And b in case {\bf PRO1}. Left: the equilibrium heating rate in Watts as a function of tidal quality factor, $Q_p$ and eccentricity. The horizontal dashed line represents $Q_p = 10^6$, the value of $Q_p$ in the right panel. Right: the equilibrium heating rate as a function of planet radius and eccentricity. Horizontal dashed lines represent the characteristic radii suggested by \cite{crossfield2010} to explain the infrared phase curve. At $i_b = 14^{\circ}$, the planet must be $1.8~R_{Jup}$ to produce the observed variational amplitude, while at $i_b = 28^{\circ}$ it must be $1.3~R_{Jup}$. The vertical gray dashed lines represent the median and peak eccentricity of the planet, $e_{med} = 0.041$ and $e_{peak} = 0.13$. The gray curve indicates the heating rate predicted by \cite{ibgui2009} for the planet HD 209458 b in the first two billion years of its tidal evolution. We find that $\upsilon$ And b can have similar internal heating rates. Heating plots for {\bf PRO2} and look very similar to those for {\bf PRO1}.}
\label{fig:heat83}
\end{figure*}

In Figure \ref{fig:heat83} we show tidal energy dissipated in the interior of planet b for case {\bf PRO1}. Tidal heating for {\bf PRO2} looks very similar.  We explore a range of tidal factor $Q$ (left panel, with planet radius $R = 1.5~R_{Jup}$) and planet radius (right panel, with $Q = 10^6$). We find that, depending on the true radius of the planet and the equation of state of the interior, planet b could indeed have episodes of intense tidal heating. Coupled with the intense stellar radiation at $\sim 0.06$ au, for the case {\bf PRO2}, this could reconcile the results of \cite{mca2010}, \cite{crossfield2010}, and this study (see the right panel in Figure \ref{fig:heat83}).  

The results are more difficult to reconcile for the lower inclination case {\bf PRO1}. \cite{miller2009} showed that radius inflation for hot Jupiters is a strong function of mass. Referring to their Figure 6, with $10^{19}$ W $= 10^{26} $ erg s$^{-1}$, and the mass of planet b of $\sim 8~M_{Jup}$ for {\bf PRO1}, it seems unlikely that even the combination of tidal heating and stellar irradiation could inflate planet b beyond $\sim 1.5~R_{Jup}$. Additionally, the planet in that case would need to have a radius of $> 3~R_{Jup}$ in order to explain the phase curve. However, given the lingering uncertainty in the apparently large radii of some hot Jupiters, as well as the shortcomings of tidal theory, these configurations may still be compatible with the \cite{crossfield2010} results.

\section{DISCUSSION}

\label{discuss}
Only four of our stable architectures (three of them prograde) are fully consistent with \cite{crossfield2010}, which requires that planet b in these cases be as large as $1.8~R_{Jup}$, and would place it amongst the largest known exoplanets. Still, this size is reasonable, as, for example, the radius of the hot Jupiter HAT-P-32 b was determined to be $R = 2.037 \pm 0.099~R_{Jup}$---a planet orbiting a star of similar spectral type (late FV) and age ($\sim 3$ Gyr) to $\upsilon$ And \citep{hartman2011}. We have demonstated that the eccentricity evolution of planet b in several cases allows for significant tidal energy dissipation which will help to explain its very large size. 
 
Yet the case with the lowest $\chi^2$ value, {\bf PRO1}, has an inclination of $\sim 5^{\circ}$ for planet b, which may be more difficult to reconcile with the \cite{crossfield2010} result. From Equation 2 in \cite{crossfield2010}, we conclude that planet b in {\bf PRO1}  would have to have a radius of $\sim 3~R_{Jup}$ to be consistent with the phase curve measurements, while its large mass may prevent inflation to even beyond $\sim 1.5~R_{Jup}$. 

There are two retrograde cases with comparable $\chi^2$ to our nominal solution, but we cautiously favor prograde orbits for planet b over retrograde on the grounds that it is difficult to explain the formation of a multi-planet system with mutual inclinations this extreme ($\Psi \sim 180^{\circ}$). Transits of many hot Jupiters have permitted detection of a Rossiter-McLaughlin effect, in which the alignment of the planet's orbit, relative to the spin axis of the star, can be inferred from asymmetries during the transit in Doppler-broadening spectral lines \citep{triaud2010}. A significant number of these orbits are misaligned with the stellar obliquity, however, as $\upsilon$ And A's obliquity is unknown, it is impossible to say at this time whether our so-called ``retrograde'' orbits are in fact misaligned with the star's rotation (or that our prograde orbit are aligned with it). Rather, it is the high mutual inclination between planet b and planets c and d that must be explained in order to justify our retrograde cases. It is noteworthy that no detected exoplanet (and host star) with $> 90^{\circ}$ misalignment has been found in a multiple planet system. Because of the lack of observational precedent, we are reluctant to favor our retrograde systems.

Early studies found that the pericenters of planets c and d were oscillating about alignment, such that $\Delta \varpi = \varpi_d - \varpi_c \sim 0^{\circ}$ \citep{chiang2001,chiang2002}. \cite{michtchenko2004} found that $\Delta \varpi$ could take on a full range of behaviors from libration to circulation, based on initial conditions within the uncertainties of the observations at the time. \cite{ford2005} found $\Delta \varpi$ near separatrix, and more recently, \cite{barnes2011} found $\Delta \varpi$ librating about anti-alignment ($\Delta \varpi \sim 180^{\circ}$). Here, we have in our prograde cases a variety of behavior for $\Delta \varpi$: precession in one case, recession in two, and libration in two.

The exact nature of the secular relationship between planets c and d appears to be highly sensitive to the orbital parameters, as we see that $\Delta \varpi$ and $\Psi_{cd}$ take on very different modes of behavior just within the observational uncertainties. It seems that the planets are close to an apsidal separatrix.  

Whereas the orbital fit \citep{mca2010} suggests that planet c has larger mass than planet d, stability seems to slightly favor planet d having the larger mass (prograde cases only), though strong conclusions should not be drawn from only four examples. While this eccentricity ratio is less likely, \cite{barnes2011} have shown that the relative inclinations are significantly more difficult to produce, and that planet-planet scattering resulting in the ejection of an additional planet is able to produce both features.  

From Figure \ref{fig:pro1}, it is clear that the system resides close to instability.  This proximity to instability and the large eccentricities and mutual inclination of planets c and d suggest that the system arrived at its current configuration by a past planet-planet scattering event, as found by \cite{barnes2011}.  The near-separatrix behavior of c and d additionally suggest that this event was more likely to be a collision than the ejection of a fourth planet from the system \citep{barnes2011}. 

\section{CONCLUSIONS}
We have presented 10 dynamically stable configurations consistent with the combined radial velocity/astrometry fit first presented in \cite{mca2010}. In six of these cases, planet b orbits retrograde with respect to planets c and d.  Because of the apparent difficulty in the formation of such a system, our analysis focuses instead on the four remaining prograde cases. The case {\bf PRO1} represents our best estimate of the system's true configuration, because of its low $\chi^2$ and the relative ease of explaining its formation.

In our stable prograde results, planet b's inclination spans the range of $3^{\circ} \lesssim i_b \lesssim 23^{\circ}$. The corresponding mass range is $1.78~M_{Jup}\leq m_b \leq 13.57~M_{Jup}$. Three of the four prograde trials are consistent with the predicted inclination from the infrared phase curve results \citep{crossfield2010}, but require a planet radius of $\sim 1.3 - 1.8~R_{Jup}$.  

$\upsilon$ Andromedae is a benchmark that may portend a new class of planetary system, i.e., ``dynamically hot'' systems with high eccentricities and high mutual inclinations.  Currently $\upsilon$ And is the only multi-planet system with astrometry measurements, but \emph{Gaia} \citep{casertano2008} and perhaps a \emph{NEAT}-like mission \citep{malbet2012} will discover if such architectures are common or rare.  If $\upsilon$ And-like systems are common in the galaxy, we should even expect to find potentially habitable planets in dynamically complex environments. 

\acknowledgments Russell Deitrick, Rory Barnes, Tom Quinn, and Rodrigo Luger acknowledge support from the NASA Astrobiology Virtual Planetary Laboratory lead team. Russell Deitrick also acknowledges support from the University of Washington GO-MAP fellowship. Rory Barnes acknowledges support from NSF grant AST-1108882. Support for HST $\upsilon$ And was provided by NASA through grants GO-09971, GO-10103, and GO-11210 from the Space Telescope Science Institute, which is operated by the Association of Universities for Research in Astronomy (AURA), Inc., under NASA contract NAS5-26555. We would additionally like to thank Jonathan Fortney, Mike Line, and Eric Agol for helpful discussions. 

\pagebreak
\clearpage

\bibliographystyle{apj}
\bibliography{ups_and_fin}

\appendix
\section{COMMENT ABOUT COORDINATES}
There is a difference in the conventions used by dynamicists and observers to relate the longitude of ascending node, $\Omega$, to the Cartesian coordinate system.  Dynamicists and dynamical models typically use the convention that $\Omega$ is measured from the $+X$-axis toward the $+Y$-axis \citep[see][]{md1999}), while observers typically measure $\Omega$ from the $+Y$-axis \citep[which typically corresponds to North, as in][]{vdk1967} toward the $+X$-axis (typically East).  

Because of this, if dynamicist's conventions are used to calculate the Cartesian coordinates of a planet's position based on the orbital elements (and these were intended for use by an observer), these coordinates would not correspond to the actual position of the planet on the sky, relative to its host star.  Rather, the $X$ and $Y$ positions would be swapped.  The reason for this can be easily understood by comparing the equations for $X$ and $Y$ from \cite{vdk1967} and those from \cite{md1999}.

\cite{vdk1967} has:
\begin{align}
X_{obs}& = B\frac{x}{a} + G\frac{y}{a} \label{xobs}\\
Y_{obs}& = A\frac{x}{a} + F\frac{y}{a}, \label{yobs}
\end{align}
where $x$ and $y$ are the positions of the planet in its own orbital plane, $X_{obs}$ and $Y_{obs}$ are the positions of the planet on the sky, using observer's conventions, and $A$, $B$, $F$, and $G$ (the Thiele-Innes constants) are: 
\begin{align}
A & = a(\cos{\omega} \cos{\Omega} - \sin{\omega} \sin{\Omega} \cos{i})\\
B & = a(\cos{\omega} \sin{\Omega} + \sin{\omega} \cos{\Omega} \cos{i}) \\
F & = a(-\sin{\omega} \cos{\Omega} - \cos{\omega} \sin{\Omega} \cos{i}) \\
G & = a(-\sin{\omega} \sin{\Omega} + \cos{\omega} \cos{\Omega} \cos{i}).\label{thieleinnesG}.
\end{align}

On the other hand, \cite{md1999} have:
\begin{align}
\begin{split}
X_{dyn}  =~ & x(\cos{\Omega}\cos{\omega}-\sin{\Omega}\sin{\omega}\cos{i}) \\
& - y(\cos{\Omega}\sin{\omega}+\sin{\Omega}\cos{\omega}\cos{i}) \label{xdyn_a}\\
\end{split}\\
\begin{split}
Y_{dyn}  =~ & x(\sin{\Omega}\cos{\omega}+\cos{\Omega}\sin{\omega}\cos{i}) \\
&- y(\sin{\Omega}\sin{\omega}+\cos{\Omega}\cos{\omega}\cos{i}), \label{ydyn_a}\\
\end{split}\\
\end{align}
where $X_{dyn}$ and $Y_{dyn}$ are the positions of the planet on the sky, using dynamicist's conventions. Equations \ref{xdyn_a} and \ref{ydyn_a} can be rewritten,
\begin{align}
X_{dyn} & = A\frac{x}{a} + F\frac{y}{a} \label{xdyn}\\
Y_{dyn} & = B\frac{x}{a} + G\frac{y}{a}. \label{ydyn}
\end{align}

\begin{figure*}
\includegraphics[width=3.6in]{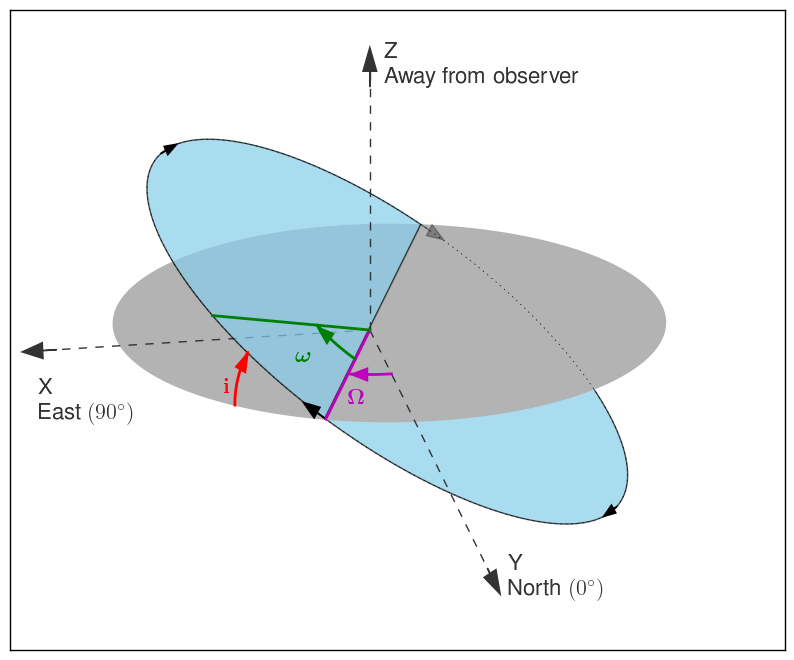}
\includegraphics[width=2.9in]{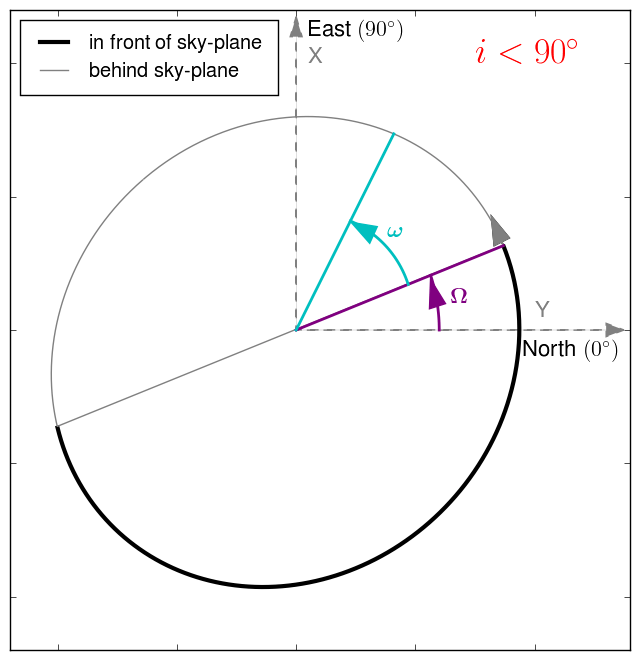}\\
\includegraphics[width=3.6in]{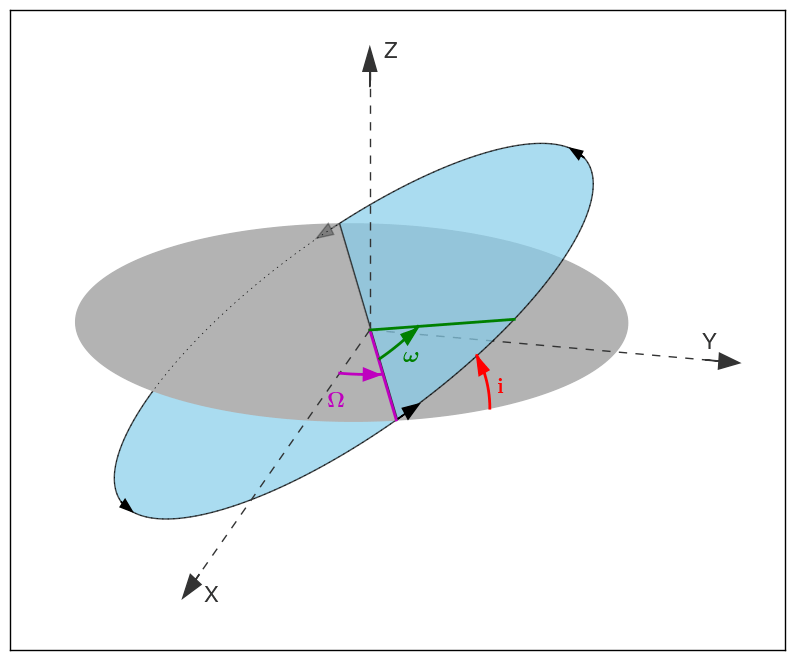}
\includegraphics[width=2.9in]{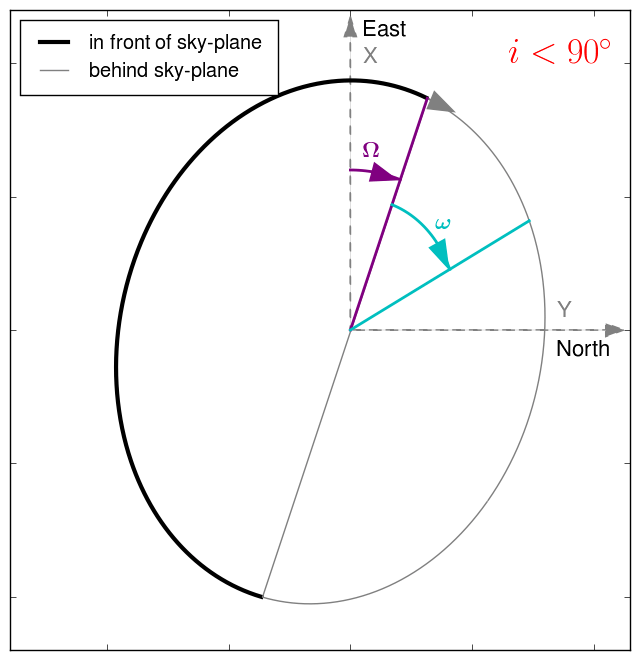}
\caption{A comparison of the orbits produced in calculating Cartesian coordinates using ``observer'' conventions and ``dynamicist'' conventions. Upper left: 3D projection of the observer's orbit, from \cite{vdk1967}. $\Omega$ is measured from the $Y-$axis (North) toward the $X-$axis (East). Upper right: The observer's orbit, projected on the sky. $\Omega$ is measured counterclockwise from the $Y-$axis (North). Lower left: 3D projection of the dynamicist's orbit. $\Omega$ is measured from the $X-$axis toward the $Y-$axis, which causes the $X$ and $Y$ coordinates of the planet to be swapped compared to the (true) observer's orbit.  Lower right: The dynamicist's orbit, projected on the sky. Again, the $X$ and $Y$ coordinates of the planet in its orbit will be swapped compared to the observer's orbit.}
\label{mirror}
\end{figure*}

Comparing Equations \ref{xdyn} and \ref{ydyn} with \ref{xobs} and \ref{yobs}, we can see that $X_{dyn} = Y_{obs}$ and $Y_{dyn} = X_{obs}$.  This means that in a dynamical model, the orbit will be a mirror image of the true observed orbit, if the observed orbital elements are taken at face value (see Figure \ref{mirror}). This point is subtle, but can lead to spurious results if an observer uses the Cartesian coordinates from an orbital simulation to plan future observations. If all observations and simulations are performed using orbital elements with their respective conventions, then all results should be consistent.

This consistency is possible because the reflection has no impact on the dynamics of the system, because all planets' orbits will be reflected about the same plane, so all relative positions and velocities are preserved, and the energy (which is a function of the semi-major axes) and angular momentum (a function of semi-major axis and eccentricity) are the same in the mirror image system as in the true system.  Thus, as long as communication of planetary properties between observers and dynamicists is restricted to Keplerian orbital elements, there should be no difficulty in correctly modeling an observed system or in making predictions of planetary positions for future observations.  

One should bear in mind the difference in Cartesian coordinates, however, when combining observational and dynamical techniques, and pass only orbital elements between models or take into account the $X/Y$ swap if necessary.

\end{document}